\documentclass[12pt]{elsart}
\usepackage{amsmath,amsfonts,amssymb,amsbsy}
\usepackage[cp1251]{inputenc}
\usepackage[english]{babel}
\usepackage{graphicx}
\usepackage{color}
\usepackage[table]{xcolor}
\usepackage{colortbl}
\definecolor{LightRed}{rgb}{1,0.8,0.8}
\begin{document}
%\setcounter{page}{1}

%\small
\begin{frontmatter}
\title{
Cylindrical lateral depth-sensing indentation testing of thin transversely isotropic elastic films:
Incompressible and weakly compressible materials}
\author{I.~Argatov},
\ead{ivan.argatov@gmail.com}
\corauth[cor]{Corresponding author.}
\author{G.~Mishuris\corauthref{cor}}
\ead{ggm@aber.ac.uk}
%\author[USA]{M.~Paukshto},
%\ead{mpaukshto@fibralignbio.com}
\address{Department of Mathematics, Institute of Mathematics, Physics and Computer Science, Aberystwyth University, Ceredigion SY23 3BZ, Wales, UK}
%\address[USA]{Fibralign Corp, 32930 Alvarado-Niles Rd., Union City, CA 94587, USA}

\begin{abstract}
An indentation testing method, which utilizes lateral contact of a long cylindrical indenter, is developed for a thin transversely isotropic incompressible elastic film deposited onto a smooth rigid substrate.
It is assumed that the material symmetry plane is orthogonal to the substrate surface, and the film thickness is small compared to the cylinder indenter length.
The presented testing methodology is based on a least squares best fit of the first-order asymptotic model to the depth-sensing indentation data for recovering three independent elastic moduli
which characterize an incompressible transversely isotropic material. The case of a weakly compressible material, which is important for biological tissues, is also discussed.
\end{abstract}

\begin{keyword}
Indentation testing \sep transversely isotropic \sep incompressible \sep thin layer \sep asymptotic model
\end{keyword}
\end{frontmatter}

\setcounter{equation}{0}
\section{Introduction}
\label{1lcSection1}

In recent years, the atomic force microscopy (AFM) indentation has become a useful technique for quantifying the mechanical properties of soft polymers \cite{Liao_et_al2010} and biological materials \cite{Stolz_et_al2009}. It is well known \cite{PandolfiVasta2012,HuangPaukshto_et_al_2013} that many biological tissues are characterized by anisotropy, and the case of transverse isotropy plays an important role for describing their symmetry properties \cite{Humphrey2003}.

Recently, the so-called cylindrical lateral indentation test, which utilizes lateral contact of a cylindrical indenter (see Fig.~\ref{figure}), was developed by
\cite{PaukshtoMishurisArgatov2014} under the assumption that both Poisson's ratios of the tested transversely isotropic elastic material are known.
Such a test can be approximately modeled \cite{ArgatovMishurisPaukshto2015} in the framework of the two-dimensional contact model for an orthotropic elastic strip, which was was studied previously in \cite{Aleksandrov2006,Erbas_et_al_2011}.
We note that in the asymptotic model \cite{ArgatovMishurisPaukshto2015} it is assumed that the film thickness $h$ and the indenter radius, $R$, should be small compared to the cylinder indenter length, $l$, as well as the contact zone size, $a$, should be small compared to the elastic film thickness $h$.

However, although it is usual to assume in the indentation testing of an isotropic material that its Poisson's ratio is known in advance \cite{ChoiZheng2005}, this assumption in the case of material anisotropy requires a careful consideration, especially for almost incompressible materials. On the other hand, in the case of incompressible transversely isotropic material, the previously developed model \cite{ArgatovMishurisPaukshto2015} needs to be further refined.

The rest of the paper is organized as follows.
In Section~\ref{1lcSection2}, the constitutive equations for a transversely isotropic material are written out in the principal coordinate system. The case of incompressible material is considered in Section~\ref{1lcSection33}.
In Section~\ref{1lcSection44}, we outline the first-order asymptotic model for the displacement-force relationship. In Section~\ref{1lcSection77}, the material identification procedure for the cylindrical lateral indentation test is outlined and the corresponding anisotropic materials properties evaluation procedure is illustrated by an example.
Finally, in Section~\ref{1lcSectionDC}, we present a discussion of the results obtained and formulate our conclusions.

\section{A transversely isotropic elastic film deposited onto a rigid substrate}
\label{1lcSection2}

We assume that the tested tissue sample forms a homogeneous, transversely isotropic, and linearly elastic layer is deposited onto a smooth rigid substrate in such a way that the plane of isotropy is perpendicular to the substrate surface.
If the $x_1 x_2$-plane coincides with the plane of isotropy, Hooke's law for a transversely isotropic material can be written as follows \cite{Lekhnitskii1981}:
\begin{eqnarray}
\varepsilon_{11} & = & \frac{1}{E}(\sigma_{11}-\nu\sigma_{22})-\frac{\nu_\Vert}{E_\Vert}\sigma_{33}, \quad
\varepsilon_{23}=\frac{1}{2G_\Vert}\sigma_{23}, \nonumber \\
\varepsilon_{22} & = & \frac{1}{E}(-\nu\sigma_{11}+\sigma_{22})-\frac{\nu_\Vert}{E_\Vert}\sigma_{33}, \quad
\varepsilon_{13}=\frac{1}{2G_\Vert}\sigma_{13}, \label{1lc(2.1)} \\
\varepsilon_{33} & = & -\frac{\nu_\Vert}{E_\Vert}(\sigma_{11}+\sigma_{22})+\frac{1}{E_\Vert}\sigma_{33}, \quad
\varepsilon_{12}=\frac{1}{2G}\sigma_{12}. \nonumber
\end{eqnarray}
Here, $\varepsilon_{ij}$ are infinitesimal strains, $\sigma_{ij}$ are stresses, $E$ and $\nu$ are Young's modulus and Poisson's ratio in the symmetry plane, $G$ is the in-plane shear modulus, $E_\Vert$ is the longitudinal Young's modulus, $\nu_\Vert$ is Poisson's ratio for loading in the longitudinal direction, and $G_\Vert$ is the longitudinal shear modulus.
Recall also that transversely isotropic materials are characterized by five independent elastic constants (namely, $E$, $E_\Vert$, $G_\Vert$, $\nu$, and $\nu_\Vert$), so that we have $G=E/[2(1+\nu)]$.

We assume that indenter has a shape of long cylinder oriented parallel to the material surface and is made from rigid material, so that the indenter deformation is negligible. Because the layer thickness is assumed to be small compared to the cylinder indenter length, we apply a two-dimensional approximation for the stress-strain state beneath the indenter.
We consider two cases, which differ by the orientation of the indenter cylinder with respect to the longitudinal axis of the transversely isotropic material. Namely, in the longitudinal case the indenter cylinder is parallel to the axis of material symmetry, while in the transverse case it is parallel to the plane of isotropy and thus perpendicular to the material symmetry axis.

\begin{figure}%[h!]
%\vskip-1.0cm
    \centering
    \includegraphics[scale=0.50]{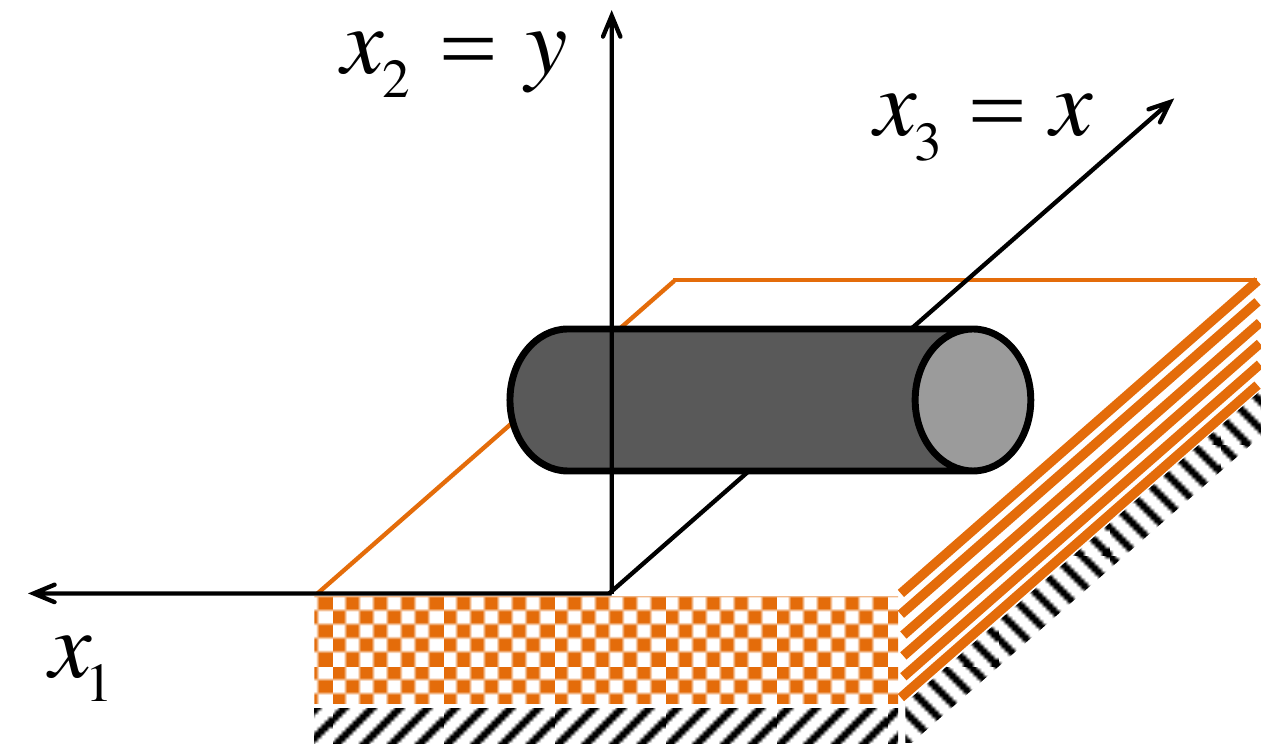}
%\vskip-3.5cm
    \caption{A homogeneous, transversely isotropic, and linearly elastic layer deposited onto a rigid substrate. Fibril direction is  perpendicular to the cylinder indenter axis.}
%\vskip-1.0cm
    \label{figure}
\end{figure}

If the indenter is oriented along the $x_1$-axis (see Fig.~\ref{figure}), the strain-stress relations for the case of plane strain deformations in the $x_2 x_3$-plane are obtained from Eqs.~(\ref{1lc(2.1)}) under the assumption that $\varepsilon_{11}=\varepsilon_{12}=\varepsilon_{13}=0$ in the following from:
\begin{eqnarray}
\varepsilon_{xx} & = & \Bigl(\frac{1}{E_\Vert}-\frac{E\nu_\Vert^2}{E^{\prime 2}}\Bigr)\sigma_{xx}-
\frac{\nu_\Vert(1+\nu)}{E_\Vert}\sigma_{yy}, \quad
\varepsilon_{xy}=\frac{1}{2G_\Vert}\sigma_{xy}, \nonumber \\
\varepsilon_{yy} & = & -\frac{\nu_\Vert(1+\nu)}{E_\Vert}\sigma_{xx}+\frac{1-\nu^2}{E}\sigma_{yy}.
\label{1lc(2.3)}
\end{eqnarray}

If the indenter is oriented along the $x_3$-axis, the strain-stress relations for the case of plane strain deformations in the $x_1 x_2$-plane are obtained from Eqs.~(\ref{1lc(2.1)}) under the assumption that $\varepsilon_{13}=\varepsilon_{23}=\varepsilon_{33}=0$ in the following from:
\begin{eqnarray}
\varepsilon_{11} & = & \Bigl(\frac{1}{E}-\frac{\nu_\Vert^2}{E_\Vert}\Bigr)\sigma_{11}-
\Bigl(\frac{\nu}{E}+\frac{\nu_\Vert^2}{E_\Vert}\Bigr)\sigma_{22}, \quad
\varepsilon_{12}=\frac{1+\nu}{E}\sigma_{12}, \nonumber \\
\varepsilon_{22} & = & -\Bigl(\frac{\nu}{E}+\frac{\nu_\Vert^2}{E_\Vert}\Bigr)\sigma_{11}+
\Bigl(\frac{1}{E}-\frac{\nu_\Vert^2}{E_\Vert}\Bigr)\sigma_{22}.
\label{1lc(2.2)}
\end{eqnarray}

Equations (\ref{1lc(2.2)}) and (\ref{1lc(2.3)}) correspond to the longitudinal (along to the
$x_3$-axis) and transverse (perpendicular to the $x_3$-axis) orientations of the cylinder indenter, respectively.

\section{Tissue incompressibility assumption and fitting parameters}
\label{1lcSection33}

If the tissue material is assumed to be incompressible, then the following condition must be satisfied: $\varepsilon_{11}+\varepsilon_{22}+\varepsilon_{33}=0$.
As was shown by \cite{ItskovAksel2002}, this incompressibility condition imposes two additional constrains on the components of the compliance tensor.
Thus, for an incompressible transversely isotropic material only 3 material constants remain independent and the following relations hold \cite{GarciaAltieroHaut1998-2,ItskovAksel2002}:
\begin{equation}
\nu_\Vert=\frac{1}{2},\quad \nu=1-\frac{E}{2E_\Vert}.
\label{1lc(2.64)}
\end{equation}
At the same time, the condition of positive definiteness of the compliance tensor reduces to
\begin{equation}
E<4E_\Vert.
\label{1lc(2.65)}
\end{equation}

Since Poisson's ratio $\nu_\Vert$ is known in advance, while Poisson's ratio $\nu$ is expressed in terms of Young's moduli $E$ and $2E_\Vert$, the problem of material parameters identification reduces to the evaluation of the following elastic constants: $E$, $E_\Vert$, and $G_\Vert$.
However, instead of these three dimensional parameters, we will make use of the transverse Young's modulus $E$ and two dimensionless ratios
\begin{equation}
n=\frac{E_\Vert}{E},\quad
k=\frac{G_\Vert}{E}.
\label{1lc(5.k)}
\end{equation}

Thus after evaluating the primary fitting elastic parameters $E$, $n$, and $k$, the elastic moduli $E_\Vert$ and $G_\Vert$ can be easily determined by formulas
\begin{equation}
E_\Vert=nE,\quad
G_\Vert=kE.
\label{1lc(5.G)}
\end{equation}

Note that in light of (\ref{1lc(2.65)}), we will have $n>1/4$.

\section{Displacement-force relation in the cylindrical lateral indentation}
\label{1lcSection44}

During the depth-sensing frictionless indentation, the indenter displacement $w$ (indentation depth) is related to the indentation load $P$ (contact force) by some nonlinear relation, which is approximated by the following first-order asymptotic formula
\cite{ArgatovMishurisPaukshto2015}:
\begin{equation}
w \simeq w_0\frac{P}{P_0} \biggl(\ln\frac{4P_0}{P}
+1-2d_0-\frac{3d_1}{2}\frac{P}{P_0}\biggl).
\label{1lc(5.1)}
\end{equation}
Here, $w_0$ is a characteristic length parameter, $P_0$ is a characteristic force parameter.

The characteristic length parameter $w_0$ is determined solely by the geometry of both the elastic sample and the rigid indenter as follows:
\begin{equation}
w_0=\frac{h^2}{4R}.
\label{1lc(5.2w)}
\end{equation}
Here, $h$ is the film sample thickness, $R$ is the radius of the indenter.

The characteristic force parameter $P_0$ is defined by the formula
\begin{equation}
P_0=\frac{\pi\theta lh^2}{2R},
\label{1lc(5.2P)}
\end{equation}
where $l$ is the length of the indenter, and depends on an elastic constant $\theta$, which will be defined later, and in the case of indentation testing, it is not known a priori.

Further, the asymptotic constants $d_0$ and $d_1$ are dimensionless parameters, which characterize not only the elastic properties of the film sample, but also the contact conditions between the sample and the rigid substrate. Under the assumption of frictionless contact between the film and the rigid substrate (i.\,e., $u_2=0$ and $\sigma_{12}=\sigma_{32}=0$ at $x_2=-h$), we can make use of the following analytical formulas \cite{Aleksandrov2006,Aleksandrov2011}:
\begin{equation}
d_0=\int\limits_0^\infty\frac{1-L(u)-e^{-u}}{u}du,\quad
d_1=-\frac{1}{2}\int\limits_0^\infty[1-L(u)]u\,du.
\label{1lc(3.1b)}
\end{equation}
Here, $L(u)$ is a dimensionless function, which depends on the elastic properties of the film sample as well as on the orientation of the indenter \cite{Aleksandrov2006}.

\subsection{Case of transverse indenter orientation}
\label{1lcSection441}

In the transverse case (indenter is oriented parallel to the $x_1$-axis and perpendicular to the $x_3$-axis), the asymptotic constants $d_0^\bot$ and $d_1^\bot$ are evaluated according to equations (\ref{1lc(5.d0)}) and (\ref{1lc(5.d1)}), where the expression for the kernel function $L(u)$ depends on the values of the elastic constants as follows
\cite{Aleksandrov2006,ArgatovMishurisPaukshto2015}:
\begin{equation}
L(u)=\left\{\begin{array}{l}\displaystyle
\frac{\kappa_1+\kappa_2}{2\kappa_1\kappa_2}L_+(u)\quad (A^2>B), \\
\displaystyle
\frac{1}{A}L_0(u)\quad (A^2=B), \\
\displaystyle
\frac{c}{c^2+d^2}L_-(u)\quad (A^2<B).
\end{array}
\right.
\label{1lc(5.7)}
\end{equation}
Here we have introduced the notation
\begin{equation}
L_+(u)=\frac{\kappa_2-\kappa_1}{\kappa_2{\,\rm cth}(\kappa_1 u)-\kappa_1{\,\rm cth}(\kappa_2 u)},\quad
\kappa_{1,2}=\sqrt{A\pm\sqrt{A^2-B}}\quad (A^2>B),
\label{1lc(2.7)}
\end{equation}
\begin{equation}
L_0(u)=\frac{{\rm ch}(2Au)-1}{{\rm sh}(2Au)+2Au},\quad
(A^2=B),
\label{1lc(2.8)}
\end{equation}
\begin{equation}
L_-(u)=\frac{d[{\rm ch}(2cu)-\cos(2du)]}{d{\,\rm sh}(2cu)+c\sin(2du)},\quad
c=\frac{\sqrt{A+\sqrt{B}}}{\sqrt{2}},\quad d=\frac{\sqrt{\sqrt{B}-A}}{\sqrt{2}} \quad
(A^2<B),
\label{1lc(2.9)}
\end{equation}

In the case of an incompressible material, the dimensionless parameters $A^\bot$ and $B^\bot$ are given by
\begin{equation}
A^\bot=\frac{2n^2-(4n-1)k}{k(4n-1)},\quad
B^\bot=1,
\label{1lc(2.5)}
\end{equation}
while the corresponding elastic constant $\theta^\bot$ is given by
\begin{equation}
\theta^\bot=\frac{2n^2 E}{4n-1}.
\label{1lc(5.d111)}
\end{equation}

We note that the notation $A^\bot$ and $B^\bot$ is consistent with the notation $E_\Vert$ and reflects the transverse orientation of the indenter cylinder to the axis of material symmetry.

\subsection{Case of longitudinal indenter orientation}
\label{1lcSection442}

In the longitudinal case (indenter is oriented along the $x_3$-axis), we have
$A^{\vert\vert}=1$ and $B^{\vert\vert}=1$. Therefore, the asymptotic constants $d_0^{\vert\vert}$ and $d_1^{\vert\vert}$ are independent of elastic properties of the sample and are evaluated as follows:
\begin{equation}
d_0^{\vert\vert}=0{.}3517,\quad d_1^{\vert\vert}=-0{.}521.
\label{1lc(5.d0)}
\end{equation}
In this case for incompressible material, the elastic constant $\theta^{\vert\vert}$ is given by formula
\begin{equation}
\theta^{\vert\vert}=\frac{2nE}{3},
\label{1lc(5.d1)}
\end{equation}
where the dimensionless parameter $n$ is defined by the first formula (\ref{1lc(5.k)}).

\section{Material identification procedure}
\label{1lcSection77}

In the transverse case, by fitting the experimental depth-load data with the analytical approximation (\ref{1lc(5.1)}) in view of (\ref{1lc(3.1b)}) and (\ref{1lc(5.7)})--(\ref{1lc(2.9)}), one can determine the elastic constants $n$, $m$, and
\begin{equation}
\theta^\bot=\frac{2R}{\pi lh^2}P_0^\bot,
\label{1lc(5.9)}
\end{equation}
where $P_0^\bot$ is the corresponding best-fit value for $P_0$ in (\ref{1lc(5.1)}).

Thus, in view of (\ref{1lc(5.d111)}) and (\ref{1lc(5.9)}), we obtain
\begin{equation}
E=\frac{4n-1}{2n^2}\theta^\bot,\quad E_\Vert=nE,\quad G_\Vert=kE.
\label{1lc(5.10)}
\end{equation}

Formulas (\ref{1lc(5.9)}) and (\ref{1lc(5.10)}) solve the formulated above material properties identification problem.

In the longitudinal case, because the asymptotic constants $d_0^{\vert\vert}$ and $d_1^{\vert\vert}$ are independent of the material properties, fitting the experimental data for the force-displacement relationship with the analytical approximation (\ref{1lc(5.1)}) allows to determine only one elastic constant~$\theta^{\vert\vert}$.

\subsection{Example. Application to indentation testing of a collagen layer}
\label{1lcSection55}

To illustrate the outlined identification procedure, we consider the same cylindrical lateral depth-sensing indentation test for a collagen coating freely deposited on a glass substrate with the fibril direction perpendicular to the cylinder axis (see Fig.~\ref{figure}) as considered in our previous study \cite{ArgatovMishurisPaukshto2015}. The collagen membrane thickness is approximately 5~$\mu\rm m$, while the indenter cylinder length is about 18~$\mu\rm m$, and the cylinder diameter is 6~$\mu\rm m$.

Under the assumption of material incompressibility, the following best-fit values were obtained:
$\theta^\bot=10{.}96$~MPa, $n=5{.}18$, $k=1{.}35$. Correspondingly, the normalized root-mean-square error of the best-fit approximation is $\delta=0{.}021$.
Now, using the first two formulas (\ref{1lc(5.10)}), we get approximation for the elastic moduli and the longitudinal shear modulus as follows: $E=4{.}03$~MPa,
$E_\Vert=20{.}86$~MPa, and $G_\Vert=5{.}4$~MPa. In light of (\ref{1lc(2.64)}) we have $\nu_\Vert=0{.}5$ and $\nu=0{.}90$. Note also that the dimensional parameter $m=G_\Vert/G$, which is related to $k$ through formula $m=2(1+\nu)k$, takes the following value: $m=5{.}13$.

Recall that under the assumption of prescribed values for the Poisson's ratios
$\nu_\Vert=0{.}45$ and $\nu=0{.}45$, the following data was obtained
\cite{ArgatovMishurisPaukshto2015}:
$\theta^\bot=10{.}60$~MPa, $n=8{.}25$, $m=1{.}05$,
$E=16{.}9$~MPa, $E_\Vert=139{.}6$~MPa, and $G_\Vert=6{.}1$~MPa, whereas $\delta=0{.}020$.
Therefore, the complete incompressibility is a rather restrictive (strong) hypothesis, since the longitudinal elastic modulus $E_\Vert$ demonstrated a 7 times decrease.
Thus, in order to be sure that the estimated Young moduli are correctly evaluated, such an assumption should be thoroughly justified and a {\it level of material incompressibility\/} should be clarified.

\section{Discussion and conclusions}
\label{1lcSectionDC}

As it follows from the previous discussion, one can expect that for weakly compressible materials the material identification data is sensitive to the choice of the Poisson's ratio values. Taking for instance the typical values $\nu_\Vert=0{.}49$ and $\nu=0{.}95$ (assumed in \cite{GilchristMurphyParnell2014} for a soft highly anisotropic tissue with $E_\Vert/E=20{.}3$) and utilizing the identification procedure based on the first-order model \cite{ArgatovMishurisPaukshto2015}, we, in particular, obtain $n=1{.}57$ and $\delta=0{.}329$ (see Table~\ref{table:nbdelta}).
Note that the best-fit approximation error turns out to be an order of magnitude higher than in the previous cases (see Section~\ref{1lcSection55}).
Moreover, the obtained value of the elastic moduli ratio $n=E_\Vert/E$ comes in contradiction with the following thermodynamic limitation imposed on the Poisson's ratio $\nu$ \cite{Lempiere1968}:
\begin{equation}
-1<\nu<1-\frac{2\nu_\Vert^2}{n}.
\label{1lc(5.Le)}
\end{equation}

\begin{table}[!h]
\caption{
Variation of the evaluated moduli ratio $n=E_\Vert/E$, the upper bound $\nu^*=1-2\nu_\Vert^2/n$ for the Poisson's ratio $\nu$, and the approximation error $\delta$ for different values of the Poisson's ratio $\nu$ and a prescribed value of the Poisson's ratio $\nu_\Vert=0{.}49$.}
\vskip0.2cm
\begin{center}
\begin{tabular}{ccccccc}\hline
$\nu$ 		&   0{.}95 		& 0{.}90 		& 0{.}85 		& 0{.}80 		& 0{.}75 		& 0{.}70  \\ \hline
$n$ 			& \cellcolor[gray]{0.9} 2{.}47 	& \cellcolor[gray]{0.9} 1{.}57 		& 13{.}52 	& 7{.}54 		& 8{.}12 		& 8{.}21 \\
$\nu^*$ 	& \cellcolor[gray]{0.9} 0{.}806 	& \cellcolor[gray]{0.9} 0{.}694 	& 0{.}964 	& 0{.}936 	& 0{.}941 	& 0{.}942 \\
$\delta$ 	& \cellcolor[gray]{0.9} 0{.}742 & \cellcolor[gray]{0.9} 0{.}329 & 0{.}139 & 0{.}021 & 0{.}020 & 0{.}023
  \\\hline
\end{tabular}
\\[4mm]
\end{center}
\label{table:nbdelta}
\end{table}

%\vspace*{5mm}

We also present in Table~\ref{table:nbdelta} the result of evaluation of the Young moduli for different values of the Poisson's ratio $\nu=0.7,0.75,0.8,0.85,0.90$. The material parameters
evaluated for $\nu=0.90$ do not also satisfy the restriction (\ref{1lc(5.Le)}) and thus cannot be accepted.

Therefore, in the case of slight incompressibility similar to that considered in \cite{GilchristMurphyParnell2014}, it is suggested first to solve the material parameters identification problem using the algorithm proposed in this paper under the assumption of the complete incompressibility, thus obtaining an upper bound for the Poisson's ratio $\nu$. After that, utilizing the algorithm \cite{ArgatovMishurisPaukshto2015}, one can adjust step-by-step the values of Poisson's ratios in such a way that the right-hand side inequality in (\ref{1lc(5.Le)}) is preserved, while the approximation error is kept to a minimum.

It is also worth emphasizing that in the case of thin layers, when the size of the contact zone is much larger than the layer thickness, the concept of a weakly incompressible layer can be introduced \cite{Mishuris2004}, which takes into account the interplay between the material and geometrical parameters. The same approach can be utilized in the indentation problem under consideration. However, in the case of anisotropic material, one should take care of two small parameters, which define how the two bulk moduli describe the material behavior when its incompressibility increases.

Summarizing, in the present paper, the first-order asymptotic model for the frictionless unilateral contact of a cylindrical indenter is applied for the development of cylindrical lateral indentation tests for thin (compared to the cylinder indenter length) transversely isotropic elastic samples under the assumption of material incompressibility. The proposed indentation material parameters identification method utilizes the depth-load indentation data for the cylinder indenter oriented orthogonal to the fiber direction only. If the depth-load data are available for the other indenter position (along the fiber direction), they can be additionally used for verifying the predicted value of the transverse modulus~$E$.

\section{Acknowledgment}

IA and GM acknowledge support from the EU projects: FP7 IRSES Marie Curie grant
TAMER No 610547 and HORIZON2020 RISE Marie Sklodowska Curie grant MATRIXASSAY
No 644175, respectively. The authors also thank M. Paukshto
for a fruitful discussion.

%\newpage

\end{document}